\documentclass[aps,prb,twocolumn,floatfix,bibliography,superscriptaddress]{revtex4-2}
\usepackage{graphics,bm,amsmath,amssymb,natbib,url,epsfig,psfrag,color}
\usepackage{graphicx}
\usepackage{subfigure}
\graphicspath{{./Figs/}{./}}

\newcommand{\be}{\begin{equation}}
\newcommand{\ee}{\end{equation}}
\newcommand{\bea}{\begin{eqnarray}}
\newcommand{\eea}{\end{eqnarray}}

\newcommand{\beq}{\begin{eqnarray}}
\newcommand{\eeq}{\end{eqnarray}}
\usepackage{xcolor}
\usepackage{amsmath}
\usepackage[colorlinks=true,pdfborder={0 0 0},linkcolor=blue]{hyperref}

\newcommand{\Eq}[1]{{\textcolor{blue}{Eq.}}~\!\!(\ref{#1})} 
\newcommand{\Fig}[1] {{\textcolor{blue}{Fig.~}}~\!\!\ref{#1}}

\begin{document}
\newcommand{\Tr}{\text{Tr}}
\newcommand{\eran}[1]{{\color{black}#1}}

\def\bs#1\es{\begin{split}#1\end{split}}	\def\bal#1\eal{\begin{align}#1\end{align}}
\newcommand{\nn}{\nonumber}
\newcommand{\sgn}{\text{sgn}}

\title{Quantum limitation on  experimental testing of non-equilibrium fluctuation theorems}
	
\author{Cheolhee Han}
\affiliation{Raymond and Beverly Sackler School of Physics and Astronomy, Tel Aviv University, Tel Aviv 69978, Israel}
\affiliation{Department of Data Information and Physics, Kongju National University, Kongju 32588, Republic of Korea}

\author{Doron Cohen}
\affiliation{Department of Physics, Ben-Gurion University of the Negev, Beer-Sheva, 84105 Israel}

\author{Eran Sela}
\affiliation{Raymond and Beverly Sackler School of Physics and Astronomy, Tel Aviv University, Tel Aviv 69978, Israel}

\date{\today}
\begin{abstract}
\eran{We consider work fluctuation theorems for isolated driven systems and the possibility to probe them in mesoscopic systems. In this context} non-equilibrium fluctuation theorems (NFTs) relate work performed on a system as its Hamiltonian varies with time, to equilibrium data of the initial and final states.  In a classical context the system energy can be directly measured, while a quantum implementation requires the incorporation of a work-agent. \eran{Here, as a work agent we consider a dynamical single-coordinate object, which exchanges energy with the system. The work done on the system is defined as the energy reduction of the work agent, which requires an energy measurement of the agent (only) at the end of the process. To justify the applicability of the NFT we require the agent's trajectory to be weakly affected by the energy exchange with the system.} We furthermore argue that the uncertainty in the energy measurements imposes an inherent quantum limitation on the validity of the NFT. We demonstrate our findings for a two-level system, and discuss  applications to more complex mesoscopic systems.    
\end{abstract}
\maketitle

\section{Introduction} 

Stochastic thermodynamics describes non-equilibrium processes of small systems governed by large fluctuations. As originally observed in classical processes, e.g. of stretching a  molecule of RNA~\cite{liphardt2002equilibrium,Collin_2005}, 
individual measurements of the work $W$
are intrinsically random. However after gathering sufficient statistics, it had been demonstrated that the nonequilibrium work distribution function (WDF), $P(W)$, satisfies general relations involving solely equilibrium free energy differences, such as the Jarzynski and Crooks relations~\cite{jarzynski1997nonequilibrium,crooks2000path}, known as non-equilibrium fluctuation theorems (NFTs)~\cite{harris2007fluctuation,esposito2009nonequilibrium,seifert2012stochastic}.

Quantum extensions of stochastic thermodynamics have been  formulated~\cite{tasaki2000jarzynski,kurchan2000quantum,mukamel2003quantum,chernyak2004effect,Huber_2008,esposito2009nonequilibrium,talkner2007fluctuation,fusco2014assessing,uzdin2015equivalence,strasberg2022quantum},   particularly,  via the  ``two-time measurement protocol"~\cite{talkner2007fluctuation,esposito2009nonequilibrium}, which incorporates   projective measurements of the energy of the system before and  after the non-equilibrium processes.

While an implementation of a projective energy measurement is possible in few-level quantum systems, see e.g. Ref.~\cite{smith2018verification}, measurement of the  energy change in many-body systems is not feasible. One common approach to overcome this problem is to measure energy {\em changes} of the system, which involves continuous monitoring of only a small sub-system. For example experiments in few-electron quantum dot (QD) systems~\cite{saira2012test,koski2013distribution,hofmann2017heat,hofmann2016equilibrium,barker2022experimental,PhysRevLett.131.220405} demonstrated NFTs, by continuously monitoring the charge state of the QD. However as in the Zeno effect, continuous monitoring leads to backaction and hence does not allow to study the WDF of quantum systems. 

Furthermore, the two-time measurement protocol goes against the ideology of thermodynamics. In a thermodynamic formulation, work $W$ and heat $Q$ are determined via measurements of external bodies. The engineer is probing the energy that is transferred to reservoirs $\mathcal{R}$ or from work agents $\mathcal{A}$, respectively, while the system itself should not be measured. 

Indeed, one can blur the distinction between work and heat, and decide to change the experimental setup such that work is converted into heat, and then a {\em calorimetric} measurement is carried out. In the proposed implementation~\cite{pekola2013calorimetric} the coupling between the quantum system and the calorimeter is the dominant energy relaxation channel, and accordingly further sophistication is required if one wants to keep the system itself coherent during the stage when work is being performed.   

Thus, verification of NFTs in general quantum systems requires a more sophisticated measurement setup.  
An inspiring  approach~\cite{PhysRevLett.110.230601,PhysRevLett.110.230602,campisi2013employing,Batalh_o_2014} uses an ancilla qubit which controls the system's Hamiltonian, and allows to extract the work statistics from the qubit's tomography. This approach is suitable to certain controllable systems like cold atoms, however it is not clear how to apply it in mesoscopic systems~\cite{saira2012test,koski2013distribution,hofmann2017heat,hofmann2016equilibrium,barker2022experimental,PhysRevLett.131.220405} in which we are interested here. 
Furthermore, it should be realized that this method does not allow direct measurement of $P(W)$, but of the associated characteristic function, which might be tricky for analysis in a practical implementation. 
%
%

Here, we study a direct approach to measure the WDF based on a ``quantum work agent"~\cite{cohen2012straightforward,monsel2018autonomous}. In this approach, we measure the energy change of a work agent. This requires the work agent to be a dynamical degree of freedom, which however introduces important limitations 
on experimental testing of NFTs.

\begin{figure}
\includegraphics[width=0.9\columnwidth]{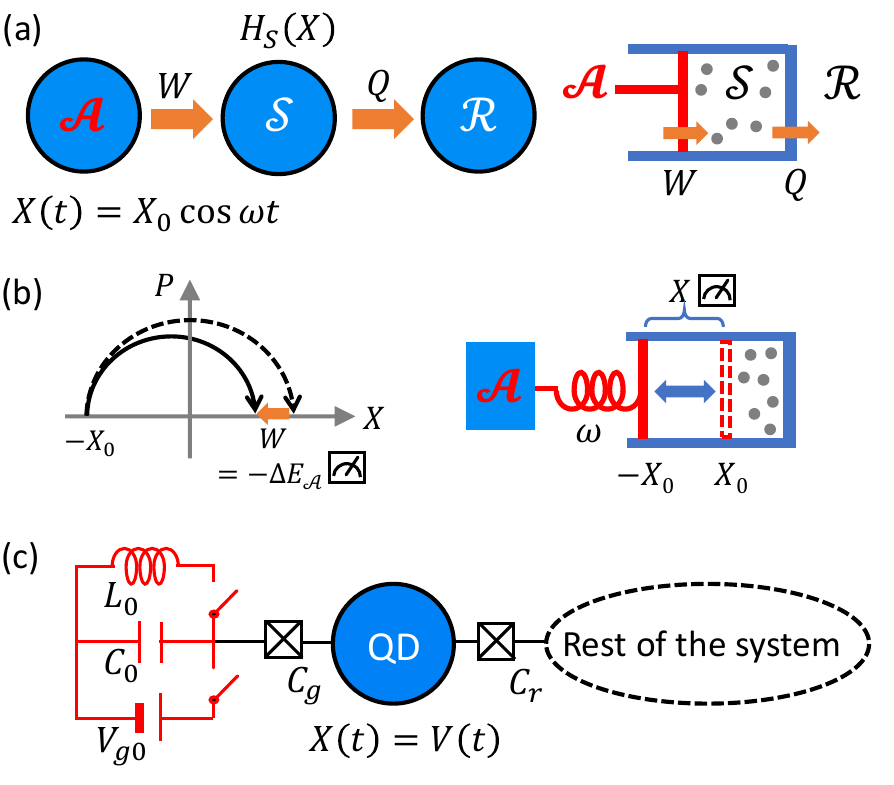}
\caption{
(a)~Thermodynamic processes driven by a work agent $\mathcal{A}$ performing work on the system $\mathcal{S}$ via a time dependent parameter $X(t)$ controlling the system's Hamiltonian. Some of it gets dissipated into a reservoir $\mathcal{R}$. 
(b)~We generalize the work agent $\mathcal{A}$ into a dynamical quantum coordinate $\hat{X}$ which performs work on the system. Its measurement 
after a half cycle allows to extract work $W=-\Delta E_\mathcal{A}$. 
(c)~Mesoscopic system: $\mathcal{S}$ contains a QD. 
Its energy level $\epsilon_d(t)$ is driven by an agent $\mathcal{A}$ realized by an $LC$ circuit such as a microcavity.
}
\label{fig:strategy}
\end{figure}

\subsection{Work agent}

\eran{The concept of a dynamical work agent is well studied in the quantum information literature in various forms. Let us review this literature and then emphasize the distinction of our approach which is motivated by experiments with mesoscopic systems. 

An intensive line of study, considers as a work agent a ``weight". The weight is an energy storage device, controlled by the gravitational potential and has an Hamiltonian being proportional to its coordinate, $H_{weight} = Mg \hat{X}$. Explicit processes have been demonstrated~\cite{skrzypczyk2014work} in which the extracted work from the system to the weight equals the free energy difference. In this line of studies, thermodynamic processes are described by a unitary evolution of the combined system, weight, and reservoir~\cite{biswas2022extraction}, and also quantum coherence in these quantum thermodynamic processes have been considered~\cite{aaberg2014catalytic,lobejko2022work,lobejko2022work}. This approach also allowed the derivation of generalized NFT theorems~\cite{alhambra2016fluctuating}. Yet, in all these studies, the weight is not dynamical, namely, the kinetic energy $\hat{P}^2/2M$ is ignored. Instead, certain quantum operations (whose experimental realization is not obvious) are routinely applied on the weight, exchanging its energy with the system~\cite{skrzypczyk2014work}. 

In another approach~\cite{beyer2020work}, work is mediated by a ``control system", which interacts unitarily with the system, and then measured, while the system itself is not measured at any time. This is another approach to avoid Zeno effects and include coherent effects in the system. In another recent study the energy storage device is referred to as a ``battery", where the interaction with the system takes the form of emission and absorption of bosons~\cite{monsel2020energetic}.

As we describe below, our approach is unique in that our work agent is meant to simulate a desired time dependent parameter $X(t)$. The work agent is a single degree of freedom such as a spring whose Hamiltonian and initial conditions are designed such that its semiclassical equations of motion coincide with the desired time dependent parameter. Yet, because it is dynamical, it exchanges energy with the system. This approach was considered~\cite{monsel2018autonomous} taking only classical effects with a continuously monitored system. Here, we consider the evolution of a closed system containing solely the system and the work agent. }

\subsection{Scope}

In the common formulation of Jarzynski’s equality the control parameter $X(t)$, for example the coordinate of a piston, is regarded as a classical coordinate, see \Fig{fig:strategy}(a).
Here we formulate a quantum demonstration of NFTs  
which incorporates a dynamical apparatus, 
referred to as a ``quantum work agent"~\cite{cohen2012straightforward}.
Of particular interest might be a sweep protocol of the form 
\beq
X(t)=-X_0 \cos(\omega t),   
\ \ \text{for} \ \  
t=0 \to \pi/\omega.
\eeq
For the purpose of an experiment we replace $X(t)$  
by a quantum dynamical coordinate $\hat{X}$, 
namely, an harmonic oscillator, see \Fig{fig:strategy}(b). 
The energy of this  single degree of freedom can be measured at $t=\pi/\omega$, independently of the complexity of the system to which it couples. 

We focus on the feasibility and limitations of the quantum work agent within a conceptual experimental realization. We characterize the ability to extract useful thermodynamic information via Jarzynski’s equality. We distinguish two effects that are ignored within traditional treatment of 
NFTs: {\bf (a)}~Quantum uncertainty of the work agent - being a quantum coordinate, the work agent yields an unavoidable uncertainty in the work measurement. 
The quantum uncertainty of the work agent  can be regarded as a variant of two-time measurement protocols with generalized measurements, including the non-demolition limit~\cite{Prasanna_Venkatesh_2014,Lahiri_2021,PhysRevE.108.024126}. 
%
{\bf (b)}~Back reaction of the system - the dynamics of the system is driven by the work agent coordinate $\hat{X}(t)$, but also affects it; 
While special attention was given to the regime of strong interaction between the system and the bath, where thermodynamic quantities  have to be properly defined~\cite{PhysRevLett.116.020601}, we emphasize that here, since the agent drives the process, its coupling to the system is naturally strong and comprises most of the energy \emph{of the system}. Thus, the energy stored in the interaction is an integral part of our definition of work. In the Born-Oppenheimer limit, where the agent is very energetic compared to the interaction energy, we show that backreaction is minimized. 
We demonstrate that one can select the parameters of the quantum work agent to minimize 
also the quantum uncertainty.

In the mesoscopic experiments in Ref.~\cite{saira2012test,koski2013distribution,hofmann2017heat,barker2022experimental,PhysRevLett.131.220405}  $X(t)$ is a time dependent gate voltage applied on a QD,  controlling its energy level and driving it across the Fermi level. To appreciate the problematic issue of continuous measurements applied in these experiments, consider e.g. a double QD coupled to a lead, where we control the gate voltage of one QD. Clearly, a continuous charge measurement of one QD  results in a suppression of the coherent oscillations within the probed system -  the quantum Zeno effect. Thus, continuous measurement destroys the quantum coherent dynamics that one would have liked to probe, which motivates  the  realization of the quantum work agent in such systems. 

As a mesoscopic realization of the quantum work agent, we propose using an $LC$ circuit, see \Fig{fig:strategy}(c), e.g. a microwave resonator. The measurement is done at the end of the process, only on the external $LC$ circuit (the ``agent"), and not on the system. 

The rest of the paper is organized as follows. After reviewing the two-time protocol to measure the WDF in Sec.~\ref{se:protocol}, we introduce our protocol based on the quantum work agent in Sec.~\ref{se:work_agenet_protocol}. In Sec.~\ref{se:general} we write down a generic model of interest for discussion of work in mesoscopic systems, and then give a general argument for the validity regime of our work agent approach. In Sec.~\ref{se:TLS} we present  a simple model of the two-level system which corresponds to a double dot system, and 
then in Sec.~\ref{se:testing} 
 we demonstrate on this model our general argument. We discuss the experimental realization of the quantum work agent in Sec.~\ref{se:exp}, and conclude in Sec.~\ref{se:summary}.

\section{The two-time protocol}
\label{se:protocol}
\label{se:2_time_protocol}

Consider an external classical parameter ${X=X(t)}$, controlling the Hamiltonian $H_{\mathcal{S}}(X)$ of a general quantum system. We assume that the system is initially at thermal equilibrium. One performs  projective energy measurements at $t=t_i$, and at the end of the process at $t=t_f$. Let $|a\rangle$ and $|b\rangle$ be eigenstates of $H_{\mathcal{S}}(X(t_i))$ and $H_{\mathcal{S}}(X(t_f))$,
with eigenvalues $E_{a}^{(i)}$ and $E_b^{(f)}$, respectively.  According to the two time protocol the WDF is defined as~\cite{talkner2007fluctuation} 
\bal
\hspace*{-2mm}
P(W)=\sum_{a,b} \frac{e^{-E_a^{(i)}/T}}{Z_i}|\langle b |U|a  \rangle|^2 \delta(W{-}(E^{(f)}_b{-}E^{(i)}_a)),\label{eq:WDF}
\eal
where 
\beq
U=\mathcal{T} \exp\left[-i\int_{t_i}^{t_f}  dt H_{\mathcal{S}}(X(t))\right] 
\eeq
is the evolution operator of the closed system and 
%
\beq
Z_{i(f)}=\Tr[e^{-H_{\mathcal{S}}(X(t_{i(f)}))/T}].
\eeq
Jarzynski's equality follows directly from this definition of $P(W)$~\cite{talkner2007fluctuation,cohen2012straightforward}, namely, 
\beq \label{eJ}
\langle e^{-W/T}\rangle_{X(t)}
&&= \int dW P(W) e^{-W/T}
\nonumber \\
&& =\frac{Z_f}{Z_i} \ \ \equiv \ \ e^{-\Delta F/T}.
\eeq
%
This identity holds for any 
system and for any non-equilibrium protocol. However, it is unpractical to perform an energy measurement of a many-body quantum system. Below, we determine the work via an energy measurement of an external work agent, which we model as a single degree of freedom oscillator. We will exclusively consider protocols of the form $X(t)=-X_0\cos(\omega t)$ from $t_i=0$ to $t_f=\pi/\omega$.

\section{The work-agent protocol} 
\label{se:work_agenet_protocol}

\eran{In this section we provide a protocol to measure work. Our guiding definition of work is the one of the two-time protocol; our subsequent protocol gives an  experimentally accessible approximation of the latter.}

The variable $X$ in reality is a dynamical coordinate of a work agent which we take to be an harmonic oscillator. The total Hamiltonian is $H=H_{\mathcal{S}}(\hat{X})+H_{\mathcal{A}}$ where
\bal
\label{eq:modelSA}
H_\mathcal{A}=\frac{\omega}{2} \left[ \left(\ell \hat{P} \right)^2 + \left( \frac{\hat{X}}{\ell} \right)^2 \right].
\eal
Both $\hat{X}$ and $\ell$ have energy units, while $[\hat{X},\hat{P}]=i$. 
It is important to emphasize that there is no ambiguity here with regard to the definition of the agent's energy. The system Hamiltonian is strictly defined as in the two-time protocol context, and includes the interaction term with $X$. The agent Hamiltonian is ``added" in a straightforward manner and does not include any system's variable.

\eran{Let us emphasize that the full Hamiltonian contains the system, the agent and the interaction between the two. We include the interaction Hamiltonian inside the system's Hamiltonain $H_{\mathcal{S}}(\hat{X})$, and use the conservation of the total energy 
\be
\label{eq:E}\frac{d}{dt} \left( \langle H_{\mathcal{S}}(\hat{X}) \rangle+ \langle H_{\mathcal{A}} \rangle  \right)=0,
\ee
to infer the work from the changes in the agent's energy. As in the two-time protocol, the total system is closed, which contains the system and the agent.}

\eran{The isolated agent's Hamiltonian has a solution which is the desired time dependent parameter whose WDF we intend to measure. In this case, this solution is $\langle \hat{X}(t)\rangle=X(t)=-X_0 \cos (\omega t)$. This can be generalized beyond the simple harmonic oscillator. If the operator $\hat{X}$ in the system's Hamiltonian could be replaced by its classical version $X(t)$, then the agent would exactly simulate the classical external parameter, and would perform exactly the desired WDF on the system.}

Our protocol is as follows:
\\ {\bf (i)}~Prepare the initial state of the agent in a coherent state at $X=-X_0$, decoupled from the system which is prepared at a thermal state according to a classical parameter $-X_0$. Namely, the initial state 
is 
\be
\rho^{(i)}=\rho_{\mathcal{S}}^{(i)} \otimes  |-X_0 \rangle \langle - X_0 | ,
\ee
where $\rho_{\mathcal{S}}^{(i)}=e^{-H_{\mathcal{S}}(-X_0)/T}/Z_i$ and $|-X_0 \rangle =e^{i \hat{P}X_0} |0\rangle$ is  a coherent state with $X_0/\ell \gg 1$.    
The initial energy of the work agent is 
$E^{(i)}_{\mathcal{A}} \approx (\omega/2) (X_0/\ell)^2$,  
where we neglect 
the numerically negligible zero point energy. 
\\ {\bf (ii)}~Let the system and agent evolve according to $H$ till $t_f=\pi/\omega$, yielding  a final state 
\be
\rho^{(f)}=e^{-iH t_f}\rho^{(i)} e^{iHt_f}.
\ee
After this process, the agent  exchanged energy and got entangled with the system.  
%
\\ {\bf (iii)}~Perform an energy measurement \emph{of the agent}, rather than  of the system. The energy measurement yields an eigen-energy $E_{\mathcal{A},n} \approx \omega n$ with probability $\rho^{(f)}_n=\Tr_\mathcal{S}[\langle n|\rho^{(f)}|n\rangle]$, where $\Tr_S[\cdots]$ is a trace over system degrees of freedom.  
We define the WDF as
\bal
P_\mathcal{A}(W) = \sum_n  \rho_n^{(f)}  \delta(W - (E_{\mathcal{A}}^{(i)} - E_{\mathcal{A},n} )).  \label{eq:WDF_a}
\eal 
We do not perform a 
measurement of the initial energy of the agent because we want it to drive the process as a coherent state, rather than a Fock state. 

\eran{From energy conservation, Eq.~(\ref{eq:E}), each result of an energy measurement of the agent yields a corresponding  energy change of the system. However, as we discuss in the next sections, the definition in Eq.~(\ref{eq:WDF_a}) does not coincide with the two-time measurement protocol due to: (i) backreaction - namely  $\hat{X}$ in $H_\mathcal{S}(\hat{X})$ does not coincide with the desired classical protocol $X(t)$;  (ii) uncertainty - even the initial energy distribution of the coherent state has a finite variance. Our goal is to formulate a validity regime in terms of the agent's parameters, where our protocol Eq.~(\ref{eq:E}) will reproduce the two-time measurement protocol. }

\section{The dot-lead system - general considerations}
\label{se:general}

\begin{figure}
\centering
\includegraphics[width=0.7\columnwidth]{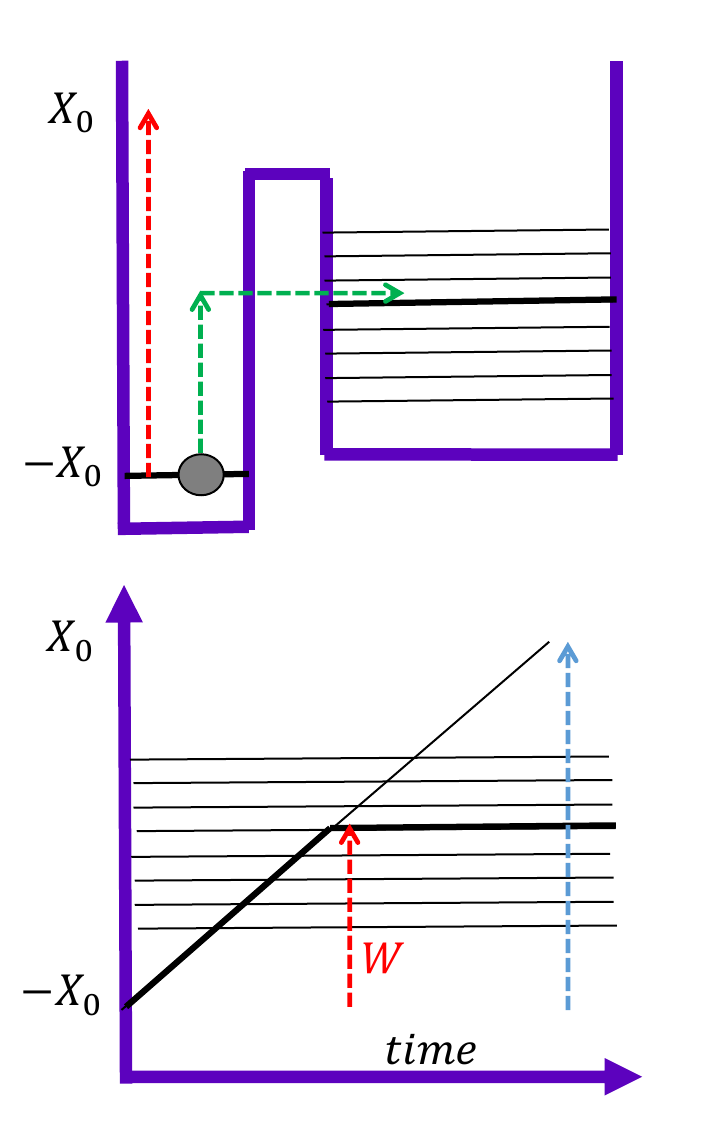}
\caption{Illustration of the multilevel dot-lead model. In the upper panel the dot level and the lead levels are represented by horizonatl balck line. The variation of the gate voltage is represented by the red arrow, and the emission of the electron from the dot to one of the lead levels by a green arrow. In the lower panel the adiabatic variation of the levels is illustrated as a function of time.}
\label{fDot}
\end{figure}

Aiming for measuring the WDF in a  general dot-lead system, let us consider a generic form of the Hamiltonian (but still within a single electron description), 
\beq
\label{eq:multilevel}
H \ &=& \  \hat{X} |0\rangle\langle 0| + \sum_{k\ne0} \mathcal{E}_k  |k\rangle  \langle k| + \ \epsilon \sum_{k\ne0} \left( |k\rangle\langle 0| + |0\rangle\langle k| \right)
\nonumber \\ 
\ &&  
+ \frac{\omega}{2} \left[ \left(\ell \hat{P} \right)^2 + \left( \frac{\hat{X}}{\ell} \right)^2 \right].  
\eeq
Here $|k \ne 0 \rangle$ denote  single particle states in the lead and $|0\rangle$ denotes the QD level. As displayed in \Fig{fDot}, regarding $\hat{X}$ as an external control parameter, the energy of the QD varies from $-X_0$ to $+X_0$. An electron originally in the QD will end up in one of the lead energy levels, or remain in the QD. This process can be viewed as a sequence of Landau-Zener transitions; the work will display a distribution function that reflects the energy at which the electron left the QD. \eran{Rather than solving for $P(W)$ with or without the agent in this model, we introduced this model in order to set the stage for the type of problems we are interested in. The aim of this section is to provide general arguments for the applicability of the protocol of Sec.~\ref{se:work_agenet_protocol}. }

Let us denote the typical (positive) work done on the system by $W_0$. We assuming that the control parameter $X$ varies from $-X_0$ to $+X_0$. Then $W_0 \sim X_0$. We do not assume here any special choice of units, neither any type of agent-system interaction. In order to minimize the backaction of the system onto the agent, it is clear that we have to require that the initial energy of the agent $\frac{\omega}{2} \left( \frac{X_0}{\ell} \right)^2$ should be much larger than $W_0$. This is essentially a Born-Oppenheimer condition,  
\beq \label{eBR}
\frac{\ell}{X_0} \ \ \ll \ \ \left(\frac{\omega}{W_0}\right)^{1/2}. 
\eeq
It should be noticed that this is a purely classical condition, that has nothing to do with quantum mechanics.
If we were using units with $\hbar\ne 1$, and then go to units with $\hbar_{\text{scaled}}=1$, then the transformation would be ${ \ell_{\text{scaled}}=\sqrt{\hbar}\ell }$ and for the energy ${ \omega_{\text{scaled}} = \hbar\omega }$. Hence the $\hbar$ dependence cancels in the above inequality.

The second  limitation  comes from the unavoidable quantum uncertainty of the energy measurement, which derives from the uncertainty in $X$. It is implied by the preparation procedure that this uncertainty is $dX\equiv\ell$. 
The associated uncertainty in the energy of the agent is 
$dE= (\omega/\ell^2)X_0 dX=\omega X_0/\ell \equiv \Delta(\ell) $. 
This uncertainty has to be much smaller than $W_0$, hence we get the {\em quantum} condition ${\Delta(\ell) \ll W_0}$.  
Combining the classical and the quantum conditions we obtain a condition for $\ell=dX$,
\beq
\label{eq:gen}
\left(\frac{\omega}{W_0}\right) \ \  \ll \ \
\frac{\ell}{X_0} \ \ \ll \ \ \left(\frac{\omega}{W_0}\right)^{1/2}.  
\eeq
In order to find $\ell$ that satisfies this double inequality the left-most expression should be smaller than the right-most expression, leading to the necessary condition
\beq
\omega \ \  \ll \ \ W_0.
\eeq
This condition means that the sweep process should be in some sense quasi-static. Clearly it is also implied that the uncertainty of the $X$ coordinate should be small:
\beq
\ell \ \ \ll \ \ X_0.
\eeq
%
%
One can rewrite \Eq{eq:gen} in terms of the range of energies that can be resolved by a given measurement setup: The energy scale $W_0$ can be regarded not just as the typical work that we want to measure, but rather as the energy scale that one wants to resolve, which might be comparable with the temperature. Accordingly we write \Eq{eq:gen} as 
\be
\omega \sqrt{N_{ph}} \ \ \ll 
\ \ \{ W_0,T \} \ \ 
\ll  \ \  \omega N_{ph},
\ee
where $N_{ph}=\frac{1}{2}\left(X_0/\ell \right)^2 \gg 1$ is the average number of photons in the oscillator (recall that here $\hbar =1$). This form of the condition stresses that the agent must be in a semiclassical state with a large number of photons.

%


\section{Two level system (TLS)}
\label{se:TLS}
\begin{figure}
\includegraphics[width=0.9\columnwidth]{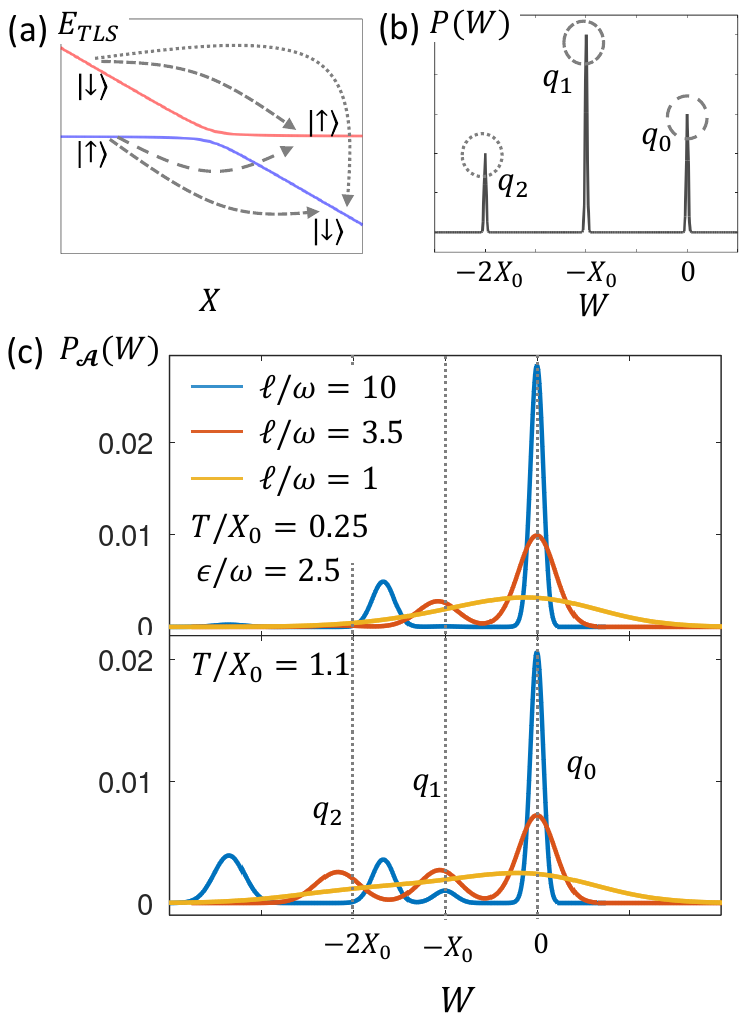}
\caption{(a) Energy levels for a TLS $\mathcal{S}$ described by Hamiltonian 
\Eq{eq:modelWOagent}. As a classical work agent $\mathcal{A}$ acts on the system by varying $X$ from $-X_0$ to $X_0$, the recorded work  gives stochastically one of the values seen in (b), corresponding to 
a diabatic transition from the thermal state ($W=-2X_0$,  dotted), adiabatic transitions ($W=-X_0$, dashed), and diabatic transition  from the ground state ($W=0$, long dashed). 
(c) The work agent is now an oscillator with energy quantization $\omega$ $(\hbar =1)$ and coordinate uncertainty $\ell$, prepared in a coherent state at position $-X_0$. 
We plot the resulting WDF $P_\mathcal{A}(W)$ according to \Eq{eq:WDF_a} for different $\ell/\omega$ and $T/X_0$ denoted with red dots in \Fig{fig:regimes1}. We discuss in the main text and in \Fig{fig:regimes1} the regimes in which $P_\mathcal{A}(W)$ gives a good approximation to $P(W)$ which allows to verify the fluctuation-dissipation theorems. 
}
\label{fig:schematics}
\end{figure}

We consider a TLS with Hamiltonian
\begin{align}
H_{\mathcal{S}}(X)=\epsilon\sigma_x+\frac{1}{2}(\sigma_z-1)X.\label{eq:modelWOagent}
\end{align}
The instantaneous ground state $|g\rangle$ and excited state $|e\rangle$ are schematically shown in \Fig{fig:schematics}(a). 
We start with a thermal state 
${\rho
^{(i)}=e^{-H_{\mathcal{S}}(-X_0)/T}}/Z_i$.

Let us first illustrate the WDF according to the two-time measurement protocol, where $X$ is a classical coordinate. The sweep of $X(t)$ induces a Landau-Zener (LZ) transition, namely, 
\beq
|g \rangle \to \sqrt{p_{d}} |e \rangle  + \sqrt{1-p_{d}} |g \rangle,
\eeq
where ${p_d=|\langle e|U|g \rangle|^2}$ is the diabatic transition probability. Specifically for $X_0\gg \epsilon$, the well known LZ formula reads ${p_d=e^{-2\pi/\alpha}}$, where ${\alpha=\omega X_0/\epsilon^2}$. Accordingly 
\bal \label{eP}
P(W) = \sum_{j=0,1,2} q_j \delta(W+j X_0),
\eal
which is illustrated in \Fig{fig:schematics}(b). 
The explicit expressions for the weights $q_j$ are
\beq
q_0 =  p_d \rho^{(i)}_{g},  \ \ && {\text{at}} \ \ W=0,
\\
q_1 = 1-p_d, \ \ && {\text{at}} \ \ W=-X_0,
\\
q_2 = p_d \rho^{(i)}_{e}, \ \ && {\text{at}} \ \ W=-2X_0,
\eeq
corresponding to the diabatic, adiabtic and thermal peaks. In these expressions  $\rho_{g(e)}^{(i)}$ is the initial probability to be in the ground (excited) state. The latter are the eigenvalues of the initial thermal state, 
\beq
\rho^{(i)} &=& \frac{1}{Z_i}e^{-H_{\mathcal{S}}(-X_0)/T}
\\
&=& \rho^{(i)}_{g} |g\rangle \langle g|
+ \rho^{(i)}_{e} |e\rangle \langle e|,
\eeq
yielding the probabilities 
\beq
\rho^{(i)}_{g,e}=e^{\pm x}/(e^x+e^{-x}),
\eeq
where 
\beq
x=\sqrt{\epsilon^2+(X_0/2)^2}.
\eeq

The diabatic peak at $W=0$ corresponds to a transition form the ground state, as opposed to the thermal peak at $W=-2X_0$ that corresponds to the diabatic transition from the thermal excited state. The adiabatic peak at $W=-X_0$ is the sum of transitions from both the ground and excited states.
In the adiabatic limit ${p_d \ll 1}$  only the adiabatic peak survives, while in the  sudden limit 
it diminishes.   
It is easily checked that \Eq{eP} satisfies Jarzynski's equality \Eq{eJ}. 
Note that for large $X_0$ we get $Z_f/Z_i \approx e^{X_0/T}$.

Results for the quantum work agent protocol are plotted in \Fig{fig:schematics}(c). Although we can identify the diabatic, adiabatic and thermal peaks of $P(W)$, these peaks have been shifted and smeared out in $P_\mathcal{A}(W)$. Below, we identify the regime within the parameter space $\{X_0, \ell,\omega,T \}$ where one can accurately use the quantum agent to verify the Jarzynski equality.

\begin{figure}
\centering
\includegraphics[width=0.45\textwidth]{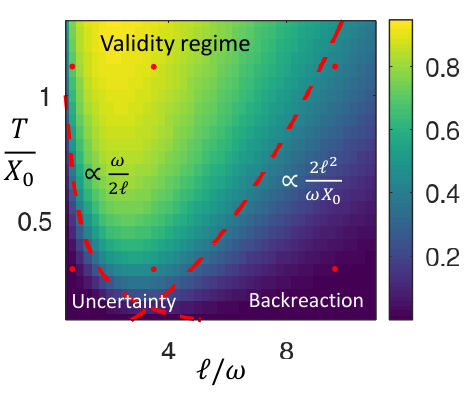}
\caption{Regime diagram: we plot the fidelity $\mathcal{F}$ of \Eq{eFid}
for $X_0/\omega=150$ and $\epsilon/\omega=2.5$.  
The red dashed curves correspond to \Eq{eq:condition_T}, which identifies the validity regime of the work agent approach. In \Fig{fig:schematics}(b) we have plotted the WDFs $P_\mathcal{A}(W)$ for the marked red dots. In the ``validity regime" both the uncertainty of the agent's coordinate, and the backreaction of the system onto the agent, are small. 
}
\label{fig:regimes1}
\end{figure}

\section{Testing the NFT with TLS} 
\label{se:testing}
The TLS example allows to test the applicability of the NFT within the quantum-work-agent framework. Assume that $P_\mathcal{A}(W)$ of \Eq{eq:WDF_a} is experimentally determined, and then used to extract the free energy via 
\beq
e^{-\Delta F'/T}=\langle e^{-W/T} \rangle_{\mathcal{A}}=\int dW P_\mathcal{A}(W) e^{-W/T}.
\eeq
The various distortions of the peaks in \Fig{fig:schematics}(c), result in $\Delta F' \ne \Delta F$. To quantify this deviation we define a fidelity measure
\beq \label{eFid}
\mathcal{F}
\equiv e^{-\frac{\Delta F'-\Delta F}{T}} 
= \left\langle  e^{-\frac{W-\Delta F}{T}}\right\rangle 
=  e^{-\frac{\delta(\ell)}{T}} \ e^{- \frac{\Delta(\ell)^2}{4 T^2}},
\eeq
where the last equality provides an analytical prediction that is explained below. This fidelity measure is plotted in \Fig{fig:regimes1}, for a given $\omega$ and $X_0$, as function of $\ell$ and $T$. It tends to unity within the validity regime. As we can see there is an intermediate regime (bright yellow region) where validity of the NFT is expected. 

The non-monotonic behavior of $\mathcal{F}$ versus $\ell$ can be understood from the combination of a broadening $\Delta(\ell)$ and shift $\delta(\ell)$ of each peak in $P(W)$. Note that the shift is also responsible for the splitting of the adiabatic peak.
For the sake of estimate we have assumed that 
\beq
\delta(W + X_0 n) \ \ \mapsto \ \ \frac{1}{\sqrt{\pi} \Delta} e^{-\frac{(W+n X_0-\delta)^2}{\Delta^2}}.
\eeq
Then we got the final expression in \Eq{eFid}.
It follows that the validity regime of the NFT is restricted by the condition
\be
\label{eq:condition_T}
\left\{ \Delta(\ell), \  \delta(\ell) \right\} \ \ \ll \ \ T .
\ee
Next we obtain the following estimates:
\be
\label{eq:delta_ell}
\Delta(\ell) \approx \frac{1}{2}\omega \frac{X_0}{\ell}
, \ \ \ \ \ \ \ \ 
\delta(\ell) \sim \frac{\ell^2}{\omega}.
\ee
%
%
The estimate for $\Delta(\ell)$ follows from the observation that there is 
an ``error" in $W$ that reflects the quantum uncertainty $\ell$ of $X$.   
In order to distinguish the peaks in \Fig{fig:schematics}(c), 
the energy uncertainty $\Delta(\ell)$ has to be smaller than~$T$. 
Irrespective of that, there is a backreaction effect that leads to the shift $\delta(\ell)$.  Also this shift should be smaller than~$T$. 
The way to obtain the estimate for $\delta(\ell)$ from conservation of energy is straightforward and is presented in the appendix.

The two inequalities of \Eq{eq:condition_T} are plotted by red dashed lines in the regime diagram in \Fig{fig:regimes1} and are highly consistent with the simulations of the protocol. 
Note that for $T \sim X_0$ the validity regime is 
\be
\label{eq:validity}
1 \ll \ell/\omega \ll \sqrt{ X_0/(2\omega) },
\ee
which is located between the region washed out by the quantum uncertainty of the energy of $\mathcal{A}$ and the region with strong backreaction in \Fig{fig:regimes1}.
This condition coincides with Eq.~(\ref{eq:gen}) which was derived independently of the details of the model.  
Within the validity regime, where the fidelity is close to unity, $P_\mathcal{A}(W)$ nicely approximates $P(W)$, see e.g. red curves in \Fig{fig:schematics}(b), where the WDF features a dominating diabatic peak at $W \sim 0$, demonstrating that the quantum work agent can operate near the sudden limit where the system is driven strongly out of equilibrium.

\section{Experimental realization}
\label{se:exp}
Consider the thermodynamic process of ramping up or down an energy level of a QD coupled to the rest of the system as in Refs.~\cite{saira2012test,koski2013distribution,hofmann2017heat,barker2022experimental,PhysRevLett.131.220405}, see \Fig{fig:strategy}(c). \eran{Rather than using the noninteracting model Eq.~(\ref{eq:multilevel}), let us consider a general Hamiltonian} 
\beq
H_{\mathcal{S}} = X(t) \hat{n}_{\text{QD}}+H_t+H_r
\eeq
where $H_t$ describes tunneling between the QD and the rest of the system which has Hamiltonian $H_r$. Our TLS example is realized if the rest of the system is another QD, 
where ${\hat{n}_{\text{QD}} = c_1^\dagger c_1}$, 
and ${H_t=\epsilon c^\dagger_1 c_2 +h.c.}$,  
while ${H_r=v c^\dagger_2 c_2}$ 
with a single electron residing in the double QD, namely $\sum_{i=1,2} c^\dagger_i c_i=1$.
We now replace the time dependent gate voltage $X(t)$ 
by a dynamical variable 
\beq
\hat{X} \ = \ \frac{e}{C_G} \hat{Q},
\eeq
where $Q=C_0V$ is the charge of a capacitor of an $LC$-circuit, 
and ${C_G = (C_g+C_r)C_0/C_g}$,  
see \Fig{fig:strategy}(c). The Hamiltonian is 
\be  
H_{\mathcal{A}} =
\left[\frac{1}{2C_0}\hat{Q}^2 \ + \frac{c^2}{2L_0}\hat{\Phi}^2 \right],
\ee
where $[\hat{Q},\hat{\Phi}]=i$. 
Comparing with \Eq{eq:modelSA} we identify
\beq
\omega &=&  \frac{1}{\sqrt{L_0 C_0}}, \\
\ell^2 &=&  e^2 \frac{C_0}{C_G^2}\omega.
\eeq
In order to probe the NFT using this $LC$-circuit 
we have to satisfy \Eq{eq:validity}, 
leading to ${\omega \ll (C_0/C_G)^2 [e^2/C_0] }$, 
and initial voltage ${V_0 \gg e/C_G}$. 

The $LC$ circuit could be a microwave resonator as in a recent experiment~\cite{ruckriegel2023dipole}.
%
Our protocol involves the following experimental challenges
: 
{\bf (i)}~prepare a coherent state in the $LC$ circuit but initially keep it decoupled from the system, which is prepared at a thermal equilibrium with fixed voltage; 
{\bf (ii)}~start ``suddenly" the sweep process by coupling $\mathcal{S}$ to $\mathcal{A}$, see switches in \Fig{fig:strategy}(c); 
and {\bf (iii)}~measure ``instantly" the energy of $\mathcal{A}$. 
In practice the switches have to be fast only compared to the typical scales of the system. For a double quantum dot this would be the maximum of the tunneling rate $\epsilon$ and dephasing rate which is on the order of GHz~\cite{ruckriegel2023dipole}. Such time control can be achieved using superconducting qubit technologies.

We may extrapolate our discussion towards exploring many body physics. 
The same agent Hamiltonian can be used for probing a general many-body quantum system. 
Considering Refs.~\cite{saira2012test,koski2013distribution,hofmann2017heat,barker2022experimental,PhysRevLett.131.220405}, the rest of the system of \Fig{fig:strategy}(c) is a spinful metallic lead with $\hat{n}_{\text{QD}}=\sum_{\sigma=\uparrow , \downarrow} d^\dagger_\sigma d_\sigma$, 
and 
\beq
H_{\mathcal{S}} &=& \ \ \epsilon_d \hat{n}_{\text{QD}}+U  d^\dagger_\uparrow d_\uparrow d^\dagger_\downarrow d_\downarrow 
\nonumber \\
&&+ \sum_{k ,\sigma} [\epsilon_k c^\dagger_{k \sigma} c_{k \sigma} + t (c^\dagger_{k \sigma} d_{ \sigma}+h.c. )].
\eeq
By coupling this Anderson model to the $LC$-circuit we can study the WDF in a process connecting two different many-body states, say, an empty state $n_{QD}=0$ with a Kondo state at $n_{QD}=1$. 
As the potential $X$ varies from $-X_0$ to $X_0$, 
an electron enters the QD at some $X$.   
With a similar reasoning as in Sec.~\ref{se:general}, as long as \Eq{eq:validity} is satisfied $P_{\mathcal{A}}(W)$ gives an increasingly good approximation for $P(W)$ even in many-body systems.

\section{Summary}
\label{se:summary}
We discussed the experimental feasibility of measuring work by employing a single-coordinate quantum object that plays the role of a ``work agent". 
To demonstrate generic aspects of testing NFTs for  quantum systems we considered a TLS toy model that can be directly realized~\cite{cottet2017observing,smith2018verification}. Then we discussed the application for many body systems of experimental interest, emphasizing that past protocols were based on continuous monitoring, hence likely to affect adversely the dynamics via e.g. a quantum Zeno effect.
Consequently, we have discussed the actual experimental realization of the simple model, as well as the potential applications for exploring many-body and Kondo physics in QDs.

There are numerous future applications of the possibility to measure $P(W)$. For example quantum critical Kondo systems~\cite{iftikhar2015two,piquard2023observing,han2022fractional} have a nontrivial WDF with Kibble–Zurek scaling~\cite{inprep2}. Also, by measuring the dissipated work one may be able to extract relative entropy~\cite{vaikuntanathan2009dissipation,deffner2010generalized,dorner2012emergent} and entanglement entropy
~\cite{ma2022symmetric,han2023realistic}.


%

\begin{acknowledgments}
   We thank Frederic Pierre, Josh Folk, Thomas Ihn, Juliette Monsel and Marcin {\L}obejko for fruitful discussions.
CH and ES gratefully acknowledge support from the European Research Council (ERC) under the European Union Horizon 2020 research and innovation programme under grant agreement No. 951541. 
DC acknowledge support by the Israel Science Foundation (Grant No.518/22).
\end{acknowledgments}

\begin{appendix}

\section{Quantum vs classical driving of the TLS}

In the two-time measurement scheme ${X(t)=-X_0\cos(\omega t)}$ is a {\em classical} control parameter that has no uncertainty nor back reaction. Below we focus numerically on two related aspects which compare this classical control parameter to the one driven by the dynamic agent.

\begin{figure}[b]
    \centering
    \includegraphics[width=0.4\textwidth]{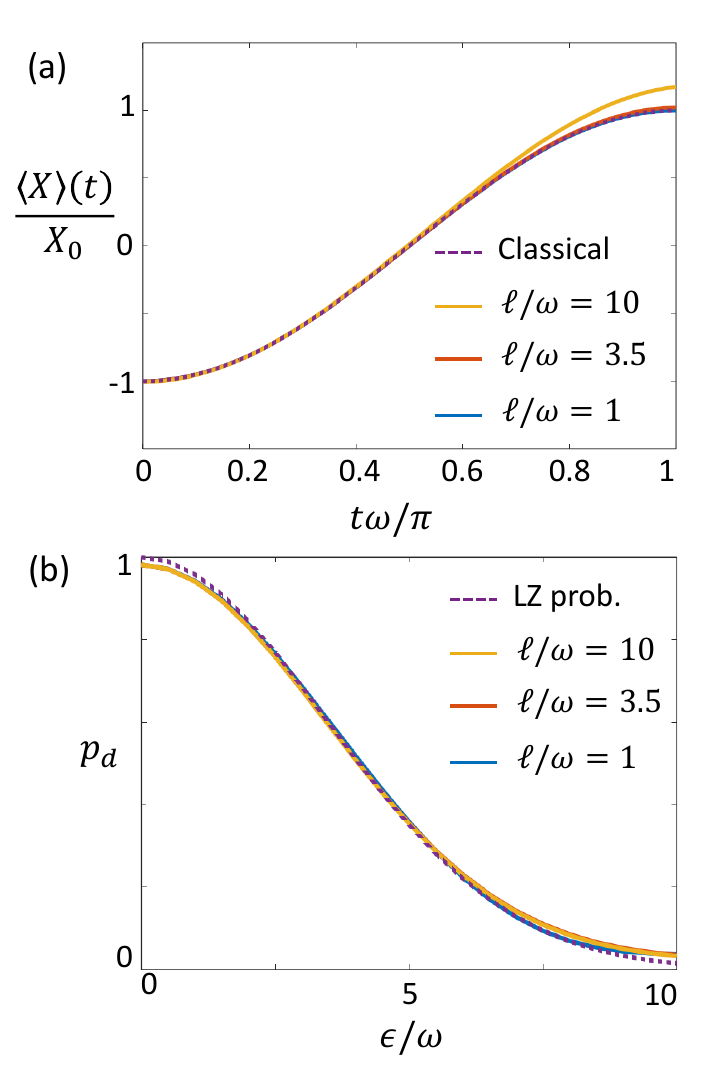}
    \caption{ (a) Time dependence of the position of the agent, for three values of $\ell/\omega$. All other parameters are the same as in Fig.~3 of the main text. As $\ell/\omega$ becomes larger, the deviation due to backreaction is larger. (b) Direct computation of the Landau-Zener transition probability, i.e. the probability to obtain spin $\downarrow$  from an initial spin-$\uparrow$ state of the TLS, after its evolution together with the agent for the process time $t=\pi/\omega$. 
} 
    \label{fig:Xt}
\end{figure}

In \Fig{fig:Xt}(a) we consider the same protocols discussed in Fig.~3 of the main text, for $T/X_0=0.25$, $X_0/\omega=150$ and three values of $\ell/\omega$.  We display the time dependence of the expectation value of the position operator of the agent, in comparison with its classical value $X(t)=-X_0 \cos(\omega t)$. As we can see,  the relative deviations become significant for the largest $\ell/\omega$. They stem from backreaction and are given by $\ell^2/(\omega X_0)$ as discussed below.

In \Fig{fig:Xt}(b) we plot  the  Landau-Zener probability $p_d$, when the process is driven by the agent. Namely, we initiate the TLS in the spin-$\uparrow$ state, and let it evolve together with the agent for the  time $t=\pi/\omega$. We define $p_d$ as the probability to find the TLS in the spin $\uparrow$ state. We are not discussing here the measurement aspect, but the agent-driven transitions versus the classically-driven transitions.  The standard Landau-Zener formula for the latter assuming linear classical driving $X(t)=X_0\omega t$ with ${t\in [-\infty,\infty]}$, yields
\beq \label{eLZ}
p_d = \exp\left[ - \frac{2\pi\epsilon^2}{\omega X_0} \right].
\eeq
As we can see, this formula works rather well even though our protocol (i) assumes sinusoidal time dependence and (ii) is driven by the quantum agent and not by a classical parameter. As $\ell$ becomes smaller, the energy uncertainty of the agent might lead to a deviation in $p_d$ that can be minimized if the average were taken over the exponent of \Eq{eLZ}, see Ref.~\cite{bourquin2012non}.


\section{Back Reaction - TLS model}


The aim of this appendix is to derive Eq.~(\ref{eq:delta_ell}) for the energy shift for the TLS. Let us  consider the adiabatic transition $\uparrow $ to $\downarrow $ where the system's energy is reduced, leading to an overshot of $X$ above $X_0$. After the transition the effective Born-Oppenheimer potential  is shifted, namely it becomes $V_{\downarrow}(X)$ instead of $V_{\uparrow}(X)$, where 
\beq
V_{\uparrow}(X) &=& \frac{\omega}{2} \left( \frac{X}{\ell} \right)^2, \\
V_{\downarrow}(X) &=& \frac{\omega}{2} \left( \frac{X}{\ell} \right)^2 - X.
\eeq  
Here, we include the interaction between the system and the agent, in the agent's potential.
The positive turning point $X'$ is implied by energy conservation 
${  V_{\downarrow}(X') = V_{\uparrow}(-X_0) }$,  
yielding 
\beq
X'=\frac{\ell^2}{\omega}+\sqrt{\left(\frac{\ell^2}{\omega} \right)^2+X_0^2}.
\eeq
Note that in the TLS-model  we have $X'>X$ because energy is taken by the agent.
By our definition the work reported by the agent is 
\beq \label{eWdef}
W=\frac{\omega}{2}\left( \frac{X_0}{\ell}\right)^2 - \frac{\omega}{2}\left( \frac{X'}{\ell}\right)^2.
\eeq
This leads to
\beq
W=
\left\{ 
\begin{matrix}  
0,  & \text{for $\uparrow$ to $\uparrow$ (diabatic transition)}     \\ 
-X', & \text{for $\uparrow$ to $\downarrow$ (adiabatic transition),}  
\end{matrix}
\right. 
\eeq
Substituting ${X' \approx X_0+\ell^2/\omega }$ we get for the adiabatic transition
\beq
W \ \ \approx \ \ \underbrace{-X_0}_{{\rm{classical}}}  \ \ \underbrace{-\ell^2/\omega}_{{\rm{back reaction}}}.
\eeq
The fact that $X'$ differs from $X_0$ is our probe as illustrated in Fig.~1(b) in the main text, but in the TLS model there is an additional adverse effect, namely, the WDF does not reproduce the exact classical work $W=-X_0$ because the classical process ${X:-X_0 \to X_0}$ is distorted. 
Consequently the energy measured by the agent is shifted 
by ${\delta(\ell) \approx \ell^2/\omega}$. This should be smaller than $X_0$, leading again to \Eq{eBR}.

\section{Back Reaction - multilevel model}

Returning to the more general dot-lead system Eq.~(\ref{eq:multilevel}), we note that the $\delta(\ell)$ complication that we have discussed for the TLS is minimized, provided that the time dependent gate affects only  the dot level, and has no effect on the potential of the lead, see \Fig{fDot}. To be specific, consider the  Hamiltonian Eq.~(\ref{eq:multilevel}). The Born-Oppenheimer potentials are  
\beq
V_{k=0}(X) 
&=& \frac{\omega}{2} \left( \frac{X}{\ell} \right)^2 + X,  
\\  
V_{k\ne0}(X) 
&=&  \frac{\omega}{2} \left( \frac{X}{\ell} \right)^2  + \mathcal{E}_k,
\eeq
where $\mathcal{E}_k$ are the levels of the lead.
In this case, for an initial state in the dot, and a final state denoted by $k$, the work done by the agent on the system is  
\beq
W=
\left\{ 
\begin{matrix}  
X_0 + X',  & k{=}0, \\ 
X_0 + \mathcal{E}_k, & k{\ne}0.   
\end{matrix}
\right. 
\eeq
%
Note that in this example the agent ``provides" work ($W>0$). Thus for a scenario where the electron does not stay in the dot, $X'$ does not affect the work, and there is no associated shift in the peak position. Indeed, in this case energy conservation $V_{k=0}(-X_0)=V_{k\ne 0}(X')$ gives 
\be
 \frac{\omega}{2} \left( \frac{X_0}{\ell} \right)^2 -X_0= \frac{\omega}{2} \left( \frac{X'}{\ell} \right)^2 + \mathcal{E}_k,
\ee
hence the work, as given by the initial minus the final energy stored in the agent, is $W=\frac{\omega}{2} \left( \frac{X_0}{\ell} \right)^2 -\frac{\omega}{2} \left( \frac{X'}{\ell} \right)^2=X_0+  \mathcal{E}_k$.
Irrespective of that, we already explained that the general condition \Eq{eq:gen} applies also to this more general model.

\end{appendix}
\bibliography{refs}

\clearpage
\end{document}